\documentclass[reprint,twocolumn,amsmath,amssymb,aps,superscriptaddress]{revtex4-2}

\usepackage{graphicx}
\usepackage{dcolumn}
\usepackage{bm}
\usepackage{graphicx}
\usepackage{amssymb}
\usepackage{epstopdf, epsfig}
\usepackage{yfonts}
\usepackage[colorlinks=true, linkcolor=blue, citecolor=blue, urlcolor=blue]{hyperref}
\usepackage{fontenc}
\usepackage{upgreek}
\usepackage{mathrsfs}
\usepackage{booktabs}
\usepackage{pifont}
\usepackage{amsmath}
\usepackage{bm}
\usepackage{wasysym}
\usepackage{caption}
\usepackage{subcaption}
\usepackage{euscript}
\usepackage{natbib}
\usepackage{lipsum}
\usepackage{xcolor}

\usepackage{orcidlink}

\usepackage{subcaption}
\usepackage{ragged2e}
\DeclareCaptionJustification{justified}{\justifying}
\newcommand{\x}[1]{\textcolor{black}{#1}}
\newcommand{\xx}[1]{\textcolor{black}{#1}}
\newcommand{\pd}[2]{\frac{\partial #1}{\partial #2}}

\newcommand{\jfm}[1]{\textcolor{blue}{#1}}

\begin{document}

\title{A Continuous Non-ergodic Theory for the Wave Set-up}

\author{Saulo Mendes\,\orcidlink{0000-0003-2395-781X}}
\email{saulo.dasilvamendes@unige.ch}
\affiliation{Group of Applied Physics, University of Geneva, Rue de l'\'{E}cole de M\'{e}decine 20, 1205 Geneva, Switzerland}
\affiliation{Institute for Environmental Sciences, University of Geneva, Boulevard Carl-Vogt 66, 1205 Geneva, Switzerland}

\begin{abstract}
Inhomogeneities in the wave field due to wave groups, currents, and shoaling among other ocean processes can affect the mean water level. In this work, the classical \xx{and unsolved} problem of \xx{continuously} computing the set-down and the following set-up induced by wave breaking on a shoal of constant \xx{finite} slope is tackled. \xx{This is possible by using available theoretical knowledge on how to} approximate the distribution of \xx{wave} random phases \x{in finite depth. Then,} \xx{the} non-homogeneous spectral analysis of the wave field \xx{allows the computation of the ensemble average by means of the phase distribution} and the inversion of the integral of the second moment for the \xx{special} case of \xx{a shoaling process with uniform phase distribution}. \xx{In doing so, I am able to} obtain a direct effect of the slope \xx{magnitude} on the phases \xx{distribution}. Therefore, \xx{an analytical and slope-dependent} mean water level with continuity over the entire range of water depth \xx{is provided}.
\end{abstract}


\maketitle

\section{INTRODUCTION}

Conservation principles in the physical sciences often have significant consequences for the solutions of governing equations \citep{Benjamin1982,Shepherd1990}. \xx{However}, discussions on the energy partition in wave hydrodynamics have risen very tardily \citep{Platzman1947,Starr1947,Whitham1962,Mack1967,Higgins1975,Klopman1990}. Not surprisingly, a strong debate surfaced over the proper formulation for the momentum of water waves \citep{Putnam1949,Eagleson1965,Bowen1969b,Higgins1970} to understand \xx{the phenomenon of} longshore currents \citep{Grant1943,Eaton1950,Caldwell1956,Inman1965}. Other nearshore phenomena were concurrently discovered \citep{Shepard1941,Shepard1950}\xx{, most prominently} the change in mean water level outside the surf zone (set-down) followed by a set-up within this zone \citep{Fairchild1958,Saville1961}. Experimental evidence \xx{supported} the radiation stress theory \citep{Higgins1962,Higgins1964} \xx{as it could seemingly} explain all these phenomena. Computation of the set-down/set-up has important consequences for adjacent water wave processes. For instance, while studying the design of breakwaters \citet{Hunt1959} showed that the run-up of waves over a beach depends on the set-up. Furthermore, rip \citep{Bowen1969,Dalrymple1978,Dalrymple2011} and longshore currents \citep{Iwata1970,McDougal1983} are also influenced by the alongshore variations in set-up. Additionally, rogue wave occurrence in the transition between deep, intermediate and shallow waters has recently been linked with the analytical effect of slope on non-homogeneous evolution of irregular waves \citep{Mendes2022b}. Remarkably, the change in mean water level can be achieved by any ocean process in which the energetic and momentum balances are disturbed, such that the classical view of the wave group-induced set-down over a flat bottom \citep{Higgins1964} is complemented by shoaling-induced \citep{Holman1985,Hsu2006b} and current-induced counterparts \citep{Brevik1978}.

The set-down is usually computed from the Bernoulli equation \citep{BMei2005}, thus the sub-harmonic is an \textit{a posteriori} term due to the conservation principles applied to a second-order irregular wave field. However, as discussed in \citet{BMei2005}, the typical calculation neglects the depth gradient, and thus overlooks the effect of slope. In addition, there is no analytical way to obtain a transition between set-down and set-up, as the latter occurs following wave dissipation and breaking which can not be applied to the Bernoulli equation. Experiments assessing the continuous evolution from wave group-induced set-down transitioning to shoaling-induced set-down until dissipation is strong, and a wave set-up appears have been conducted since the 1960s \citep{Bowen1968}. Despite the elapse of half a century, no continuous theoretical model has been developed to reproduce such an evolution other than the piecewise formulation based on the radiation stress \citep{Higgins1962,Higgins1964}. Indeed, \citet{Battjes1974} provided the best empirically-driven closed-form model for the set-up, but does not overcome the piecewise formulation between set-down and set-up. Likewise, although \citet{Hsu2006} generalized the set-down and set-up calculations of \citet{McDougal1983} for oblique waves under the effect of refraction and \citet{Gourlay2000} generalized the problem to include bottom friction, these two examples show that theoretical and numerical advances have still not been able to overcome the piecewise decomposition. 

The field of wave statistics is often treated as a mere consequence of the solutions to the governing equations of hydrodynamics, and fundamental ocean processes are not believed to be affected by extreme wave occurrence or joint probability densities of surface elevation and random phases. In this work, I follow the footsteps of \citet{Mendes2023} and demonstrate that wave statistics can indeed affect the fundamental physical process leading to the set-down and set-up, whose dynamical evolution can not be treated by a unique approach due to wave breaking.

\section{Governing Equations and Statement of the Problem}\label{sec:set0}

Given the exact solution \xx{of} the generalized velocity potential $\Phi (x,z,t)$ and surface elevation $\zeta (x,t)$ one can compute integral quantities and their conservation such as mass, momentum or energy flux \citep{Whitham1962,Higgins1975}.
Solving the Bernoulli equation leads to a clear dependence of the set-down on the mathematical form of the solution for the velocity potential \citep{BMei2005}:
\begin{equation}
 \langle \zeta \rangle = -\frac{1}{2g}   \left[ \left\langle \left( \frac{\partial \Phi}{\partial x}  \right)^{2}  \right\rangle -  \left\langle \left( \frac{\partial \Phi}{\partial z}  \right)^{2}  \right\rangle   \right] \equiv - \frac{ \langle u^{2} \rangle - \langle w^{2} \rangle }{2g}  \, .
 \label{eq:Bernoulli}
\end{equation}
 The velocity potential up to second-order in steepness can be written as \citep{Dalrymple984,Dingemans1997}:
\begin{equation}
\Phi  = \frac{a\omega}{k} \frac{\cosh{\theta} }{\sinh{\Lambda}} \sin{\phi} + \left(  \frac{3ka}{8}  \right) \frac{a\omega}{k} \frac{\cosh{(2\theta)} }{\sinh^{4}{\Lambda}} \sin{(2\phi)} \quad,
\end{equation}
with notation $\theta = k (z+h)$, $\Lambda = kh$ and $\phi = kx - \omega t$. If $\nabla h \equiv \partial h / \partial x = 0$ is assumed, the horizontal component of the velocity vector \xx{reads} ($\partial \phi/\partial x = k$\xx{, $\sin_{x} {\phi} \equiv \partial [\sin{\phi}]/\partial x$}): 
\begin{eqnarray}
\hspace{-0.2cm}
\nonumber
u &=&  \frac{a\omega}{k} \Bigg\{ \frac{\cosh{\theta} }{\sinh{\Lambda}} \xx{\sin_{x}{\phi}}  
+ \left(  \frac{3ka}{8}  \right) \frac{\cosh{(2\theta)} }{\sinh^{4}{\Lambda}} \xx{\sin_{x}{(2\phi)}} \Bigg\} \, ,
\\
&=& a\omega \left\{ \frac{\cosh{\theta} }{\sinh{\Lambda}} \cos{\phi}  + \left(  \frac{3ka}{4}  \right) \frac{\cosh{(2\theta)} }{\sinh^{4}{\Lambda}} \cos{(2\phi)} \right\} .
\label{eq:U}
\end{eqnarray}
Likewise, noting that $\partial h / \partial z = 0$ by definition and $\partial \theta/\partial z = k$, the vertical component of the velocity reads:
\begin{eqnarray}
\nonumber
w &=&  \frac{a\omega}{k} \Bigg\{ \frac{ \sin{\phi}}{\sinh{\Lambda}} \xx{ \cosh_{z}{\theta} }  
 + \left(  \frac{3ka}{8}  \right) \frac{ \sin{(2\phi)} }{\sinh^{4}{\Lambda}} \xx{ \cosh_{z}{(2\theta )} } \Bigg\} \, ,
\\
&=& a\omega \left\{ \frac{\sinh{\theta} }{\sinh{\Lambda}} \sin{\phi}  + \left(  \frac{3ka}{4}  \right) \frac{\sinh{(2\theta)} }{\sinh^{4}{\Lambda}} \sin{(2\phi)} \right\} \, .
\label{eq:W}
\end{eqnarray}
For the next step of taking the time average of the square of the velocity components, the reader \xx{is} reminded of periodic averaging of trigonometric functions (see eq.~(\ref{eq:ensembletimeave})):
\begin{equation}
 \lim_{T \rightarrow +\infty}  \int_{0}^{T} \sin^{2n+1}{\phi} \, \frac{dt}{T} = \lim_{T \rightarrow +\infty}  \int_{0}^{T} \cos^{2n+1}{\phi} \, \frac{dt}{T} = 0  \, ,
\end{equation}
\xx{for all $n \in \mathbb{N}^{\ast}$.} Through integration by parts, one has as a corollary for all $(m,n) \in \mathbb{N}^{\ast}$:
\begin{eqnarray}
\langle \sin^{2n+1}{\phi} \, \cos^{2m+1}{\phi} \rangle &=&  \langle \sin^{2n}{\phi} \, \cos^{2m+1}{\phi} \rangle  =  0 \, ,
\label{eq:sincos}
\end{eqnarray}
where the operator $\langle \cdot \rangle$ denotes time averaging. The square of the velocity components will have only two non-vanishing terms $\langle \sin^{2}{\phi} \rangle = \langle \cos^{2}{\phi} \rangle = 1/2$. \xx{Therefore,}
by means of eqs.~(\ref{eq:Bernoulli},  \ref{eq:U}-\ref{eq:W}) the set-down may be computed over a flat bottom:
\begin{widetext}
\begin{eqnarray}
\nonumber
\langle \zeta \rangle &=& -\frac{(a\omega )^{2}}{2g} \Bigg\{   \frac{\Big[\cosh^{2}{\theta} \langle \cos^{2}{\phi} \rangle - \sinh^{2}{\theta} \langle \sin^{2}{\phi} \rangle \Big]}{\sinh^{2}{\Lambda}} 
+ \left(  \frac{3ka}{4}  \right)^{2} \frac{\Big[\cosh^{2}{(2\theta)} \langle \cos^{2}{(2\phi)} \rangle - \sinh^{2}{(2\theta)} \langle \sin^{2}{(2\phi)} \rangle \Big] }{\sinh^{8}{\Lambda}} \Bigg\} \quad ,
\\
&=& -\frac{a^{2}}{2g} \cdot \frac{\omega^{2}}{2\sinh^{2}{\Lambda}} \Bigg\{   \cosh^{2}{\theta} - \sinh^{2}{\theta}  + \left(  \frac{3ka}{4}  \right)^{2} \frac{\Big[ \cosh^{2}{(2\theta)} - \sinh^{2}{(2\theta)} \Big] }{\sinh^{6}{\Lambda}}   \Bigg\} \, \quad ,
\end{eqnarray}
\end{widetext}
which taking into account the hyperbolic identity $\cosh^{2}{\theta} - \sinh^{2}{\theta} = 1$ and the leading order in the dispersion ($\omega^{2}=gk \tanh \Lambda$) leads to the formula:
\begin{equation}
\langle \zeta \rangle \approx - \frac{ka^{2}}{2\sinh{(2kh)}} \left[ 1 + \frac{9 (ka)^{2}}{16 \sinh^{6}{kh}}    \right] \quad .
\label{eq:setdownRS}
\end{equation}
As far as \xx{the author is} aware, the term of second order in steepness inside the brackets does not appear in the literature, likely being neglected without further discussion. In the limit of second order theory through the Ursell number $\textrm{Ur} \leqslant 8\pi^{2}/3$ \citep{Dalrymple1978}, the term inside brackets increases the overall set-down by not more than 15\%. However, beyond the limit of the  second-order theory, the additional term will cause the set-down to diverge. Thus, this is a symptom of the unsuitability of this approach in computing the set-down/set-up over continuous range in relative water depth. 

Alternatively, one may compute the gradient of the set-down from the horizontal gradient of the radiation stress, as the momentum balance equations lead to the radiation stress formula \citep{Higgins1962,Higgins1964}:
\begin{equation}
 \nabla S_{xx}  = - \rho g (\langle \zeta \rangle + h) \nabla \langle \zeta \rangle     \quad ,  
\label{eq:radstress}
\end{equation}
where the cross-shore component of the radiation stress is a function of the ratio between phase and group velocities:
\begin{equation}
S_{xx} = \left\langle \int_{-h}^{\zeta} (p + \rho u^{2}) \, dz     \right\rangle -  \int_{-h}^{\zeta} p_{0} \, dz \,\, , 
\end{equation}
\xx{where $p = p_{0} + \rho g \zeta = \rho g (\zeta - z)$ is the pressure field underneath the waves.}
With a little algebra, \xx{limited to the first order in steepness} this integration \xx{leads to}:
\begin{eqnarray}
S_{xx} &=& \rho \, \Bigg\langle  \int_{-h}^{\zeta} ( u^{2} - w^{2} ) dz \Bigg\rangle + \frac{1}{2} \rho g \langle \zeta^{2} \rangle  \,\, ,
\\
\nonumber
&=&  \frac{1}{2} \rho g \langle \zeta^{2} \rangle - 2 \rho g h  \langle \zeta \rangle =   \frac{1}{2} \rho g a^{2} \left[ \frac{1}{2} + \frac{2kh}{\sinh{(2kh)}} \right]   \, .  
\end{eqnarray}
This procedure verifies that eq.~(\ref{eq:setdownRS}) up to the first order is the solution of eq.~(\ref{eq:radstress}). Hence, there is no advantage in integrating the latter equation as compared to perform derivatives in eq.~(\ref{eq:Bernoulli}). Furthermore, the set-down computed above was derived without any need for shoaling formulae, which suggests there is little difference between wave-group set-down driven by difference in wave heights and shoaling of a very mild slope $\nabla h \lesssim 1/100$. Noteworthy, waves may break in the region of validity of second-order waves, and as soon as the waves reach the surf zone and are subject to strong dissipation the solution to eq.~(\ref{eq:radstress}) is no longer the equivalent of solving Bernoulli's equation. Nonetheless, the set-down caused by the cross-shore component of the radiation stress can not be properly formulated in the surf zone even in eq.~(\ref{eq:radstress}), except if one assumes that the wave height stays constant within this region \citep{Higgins1964,Bowen1968,Battjes1974}. The major issue tackled by this work is to find a continuous analytical way to express the mean water level in these two physically different zones. This can not be done either through eq.~(\ref{eq:radstress}) nor eq.~(\ref{eq:Bernoulli}). Hence, waves of second-order in steepness follow an approximated piecewise formula for the set-down and set-up, see for instance \citet{Hsu2006b}. In the next sections, I attempt to describe both set-down and set-up continuously over a plane beach within a single physical principle applicable to both zones.

\section{Random phase distribution and Wave Statistics}

Although wave dissipation typically separates physical theories of ocean wave mechanics between prior to breaking and after breaking, statistical measures of irregular waves do not suffer from this dynamical problem, at least empirically. For instance, \citet{Glukhovskii1966} showed that is possible to connect in a continuous range for relative water depth ($k_{p}h$) in terms of the ratio $H_{s}/h$ the distributions of wave heights in deep and shallow water, otherwise known to be restricted to piecewise solutions (See \citet{Ewans2016}, \citet{Mendes2021c} and \citet{Karmpadakis2022} for distributions dependent on this ratio). The latter ratio is often found to be a good proxy for wave breaking \citep{Battjes1974,Hallowell2015}. As such, in this section I invoke the practical knowledge of wave statistics and reveal how the non-ergodicity of irregular water waves plays a role in oscillating the mean water level from its initial condition (without waves) continuously as they approach a plane beach. Let the ensemble average of a random variable $X(t)$ \xx{be defined}:
\begin{equation}
 \mathbb{E}\left[ X(t) \right] = \int_{0}^{+\infty} X(t) \, f(X) \, dX 
\quad , 
\label{eq:ensembletimeave}
\end{equation}
where $f(X)$ is the probability density of the random variable $X(t)$. Then, a stochastic process is said to be \textit{ergodic} if $\mathbb{E}\left[ X(t) \right] = \langle X(t) \rangle$ holds. Now, it is of interest to compute ensemble averages of powers of the surface elevation $\zeta^{n}$ for $n \in \mathbb{N}^{\ast}$. The mean water level can be easily computed in the case of linear waves \citep{Airy1845}, and it will always leads to an ergodic process over a flat bottom with $\mathbb{E}[\zeta]=\langle \zeta \rangle=0$. Through a change of variables \citep{Papoulis2002} and applying the law of the unconscious statistician \citep{Hwang2019} to eq.~(\ref{eq:ensembletimeave}), I compute the ensemble average as follows:
\begin{eqnarray}
 \mathbb{E}\left[ \zeta \right] = \int_{-\infty}^{+\infty} \zeta \, f(\zeta) \, d\zeta = \int_{0}^{2\pi} \zeta (\phi) f(\phi) \, d\phi \quad . 
\label{eq:ensembletheta1}
\end{eqnarray}
Eq.~(\ref{eq:ensembletheta1}) delineates how the probability density of the surface elevation is transformed into the distribution of random phases in computing the very same ensemble average. In the case of linear waves, ergodicity is corollary of an uniform distribution of phases (or alternatively, a Gaussian distribution of the surface elevation):
\begin{eqnarray}
\nonumber
 \mathbb{E}\left[ \zeta \right] &=& \sum_{i}   \frac{a_{i}}{2\pi} \int_{0}^{2\pi} \cos \phi \, d\phi = \langle \zeta (t) \rangle \, ,  
 \\
 &=&  \lim_{T \rightarrow +\infty} \sum_{i} \frac{a_{i}}{T} \int_{0}^{T} \cos{\phi}  \, dt = 0   \,\, .
\label{eq:erg1}
\end{eqnarray}   
Likewise, the second moment (the spectral energy density) also features ergodicity:
\begin{eqnarray}
 \mathbb{E}\left[ \zeta^{2} \right] &=& \sum_{i}   \frac{a^{2}_{i}}{2\pi} \int_{0}^{2\pi} \cos^{2} \phi \, d\phi = \langle \zeta^{2} (t) \rangle  \, ,
 \\
 \nonumber
 &=&  \lim_{T \rightarrow +\infty} \sum_{i} \frac{a^{2}_{i}}{T} \int_{0}^{T} \cos^{2}{(\omega_{i}t)}  \, dt = \sum_{i} \frac{a^{2}_{i}}{2}   \,\, , 
\label{eq:erg2}
\end{eqnarray}
which is well-known to be related to the \citet{Khinchin1934} theorem. Consequently, \xx{it can be demonstrated as in \jfm{appendix} \ref{sec:Rayapp} that} any \xx{signal which} can be described as a two-dimensional random walk, is zero mean with finite variance and independent random variables, will result in a Rayleigh distribution \citep{Leadbetter1967,Vanmarcke2010}. \xx{For a review,} see section 7.7-2 of \citet{Middleton1996} \xx{and 6.2.2 of \citet{BMendes2020}.} 
\xx{However,} if the distribution of random phases is not uniform the envelope will deviate from the Rayleigh law. Indeed, when waves travel over a shoal the distribution of surface elevation, crests or crest-to-trough heights are strongly deviated from the normal law \citep{Trulsen2020,Chabchoub2019,Mori2021}. The latter can be understood as random phases no longer being uniformly distributed, as demonstrated by \citet{Bitner1980}. The above statistical formulation computes the mean water level from the time average \xx{as} if the \xx{shoaling} process \xx{was} ergodic. \xx{Although the latter is not factual, for a variety of purposes the ergodic approximation is useful and greatly simplifies calculations. Compared to the deterministic approach of \jfm{section} \ref{sec:set0},} the \xx{possible generalization of the} statistical approach \xx{in eq.~(\ref{eq:erg1})} has the advantage \xx{of continuously assessing} the mean water level in the surf zone.

\section{Non-ergodic Continuous Set-up}\label{sec:setdown}

\xx{In this section I attempt} to compute the ensemble average from \xx{the} random phase \xx{approach. Consider} the case of spatial inhomogeneity due to a shoal. In this case, the time series is approximately weakly stationary. The evolution of the non-homogeneous spectrum is better formulated as a homogeneous spectrum corrected by a term $\Gamma = \langle \zeta^{2} \rangle / \mathscr{E}$ that absorbs such inhomogeneity, where $\mathscr{E}$ is the spectral energy density factoring out the linear energy $(1/2) \rho g a^{2}$. Up to second order in steepness, \xx{one} would find a surface elevation in its simplest form \citep{Dingemans1997}:
\begin{equation}
 \zeta  = \sum_{i} a_{i} \cos{\phi} + \sum_{i}  a_{i}  (ka)_{i}  \left[ \frac{3 - \tanh^{2}{(k_{i}h)} }{ 4\tanh^{3}{(k_{i}h)}  } \right] \cos{(2\phi)}   \, .
 \label{eq:zetairregular}
\end{equation}
In the notation of \citet{Mendes2022b} I can rewrite it as ($\varepsilon_{i}=2a_{i}/\lambda_{i}$),
\begin{figure*}[tbh!]
\centering
    \includegraphics[scale=0.58]{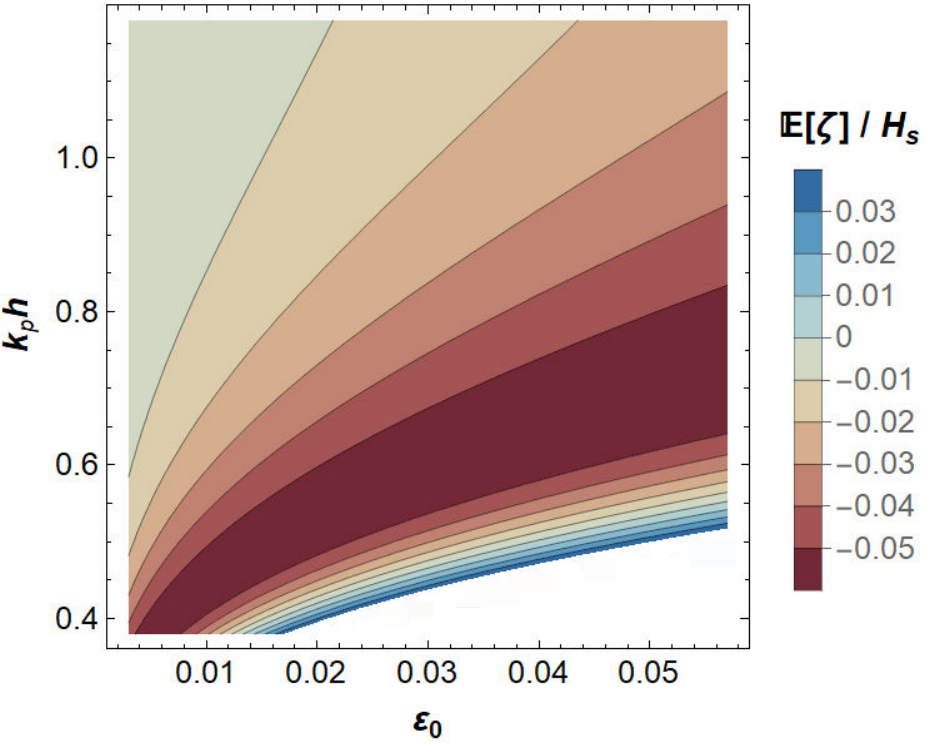}
\caption{Contour of the normalized set-down/up as a function of pre-shoal steepness $\varepsilon_{0}$.}
\label{fig:moment11}
\end{figure*}
\begin{equation}
 \zeta (x,t) = \sum_{i} a_{i} \cos{\phi} + \sum_{i}  a_{i} \cos{(2\phi)}   \left( \frac{ \pi \varepsilon_{i} \sqrt{\tilde{\chi}_{1}} }{ 4  } \right)    \, .      
\end{equation}
The $\Gamma$ correction simply reads \citep{Mendes2021b}:
\begin{eqnarray}
\Gamma &=& \frac{\mu_{2}}{\mathscr{E}} = \frac{ 1 + \left( \frac{\pi \varepsilon}{4} \right)^{2} \tilde{\chi}_{1} }{ 1 + \left( \frac{\pi \varepsilon}{4} \right)^{2} \left( \frac{ \tilde{\chi}_{1} + \chi_{1} }{2}\right) }  \quad  ,
\\
\nonumber
\Tilde{\chi}_{1} &=&  \left[ \frac{3 - \tanh^{2}{(k_{p}h)} }{ \tanh^{3}{(k_{p}h)}  } \right]^{2}  \,\, , \,\,  \chi_{1} = \frac{9\, \textrm{cosh}(2k_{p}h) }{\textrm{sinh}^{6} (k_{p}h)} \, ,
\label{eq:Gamma}
\end{eqnarray}
where $\varepsilon = H_{s}/\lambda$ denotes the irregular wave measure of significant steepness, with $H_{s}$ being the significant wave height and $\lambda$ the zero-crossing period. The effect  of the $\Gamma$ correction on the moments of $\zeta$ in a uniform distribution of phases \xx{up to second order follows} (note that the Airy case has $\mathbb
{E}[\zeta^{2}] = 1)$:
\begin{eqnarray}
\nonumber
\mu_{1} &=& \frac{\mathbb{E}[\zeta]}{\sqrt{\mu_{2} }} = \frac{\sqrt{2}}{a} \frac{1}{\sqrt{1 + \left( \frac{\pi \varepsilon}{4} \right)^{2} \tilde{\chi}_{1}}} \xx{\times}
\\
\nonumber
&{}&\int_{0}^{2\pi} \frac{1}{2\pi} \left[a \cos{\phi} + \frac{\pi \varepsilon}{4} \sqrt{\tilde{\chi}_{1}} \cdot a \cos{(2\phi)} \right]  \, d\phi = 0 \quad ,
\\
\nonumber
\mu_{2} &=& \frac{2}{a^{2}} \int_{0}^{2\pi} \frac{1}{2\pi} \left[ a \cos{\phi} + \frac{\pi \varepsilon}{4} \sqrt{\tilde{\chi}_{1}} \cdot a \cos{(2\phi)} \right]^{2}  \, d\phi
\\
&=& 1 + \left( \frac{\pi \varepsilon}{4} \right)^{2} \tilde{\chi}_{1} \quad .
\label{eq:randomphaseGamma}
\end{eqnarray}
If linear wave theory is assumed ($\zeta = a \cos{\phi}$) but with slightly non-uniform phase distribution affected by $2 \pi f(\phi) = 1 - \frac{\pi \varepsilon}{4} \sqrt{\tilde{\chi}_{1}} \, \cos{\phi}$, $\mu_{2}$ is recovered to second order. This suggests one can estimate the magnitude of the perturbation over the otherwise uniform distribution $2 \pi f(\phi) - 1 \sim - \frac{\pi \varepsilon}{4}  \sqrt{\tilde{\chi}_{1}} \, \cos{\phi} $. Therefore, a generalized distribution of the form is sought:
\begin{equation}
2\pi f(\phi) = \left[  1 - \left( \frac{\pi \varepsilon}{4}  \sqrt{\tilde{\chi}_{1}}  \right) \frac{\cos{\phi}}{\Xi_{1}} + \left( \frac{\pi \varepsilon}{4}  \sqrt{\tilde{\chi}_{1}}  \right)^{2} \frac{\cos{(2\phi)}}{\Xi_{2}} \right] \, ,
\label{eq:disthyp}
\end{equation}
where $(\Xi_{1},\Xi_{2})$ are coefficients to be \xx{uniquely} determined by wave theories. \xx{As a remark}, from the point of view of the probability distribution \citep{Mendes2021b} the second moment \xx{has been} normalized by the energy, now modified by a second-order correction. In this case,  the true second \xx{normalized} moment is $\mu_{2}/\mathscr{E} = \Gamma$. Likewise, the first moment reads  $\mu_{1} = \mathbb{E}[\zeta]/\sqrt{\mathscr{E}\mu_{2}}$.

\subsection{Computation through Gram-Charlier Series}

To study non-uniform distributions of random phases one must obtain the joint probability density of both phases and envelope (the analytical equivalent of the wave height). The non-uniformity of random phases is therefore closely related to the non-Gaussianity of the surface elevation probability density. Historically, deviations from Gaussian law are dealt through asymptotic expansions of the \textit{central limit theorem}, having been rigorously defined since \citet{Laplace1812} with further refinements introduced by \citet{Chebyshev1887}, \citet{Berry1941} and \citet{Esseen1942}. The approximation through Gram-Charlier or Edgeworth series of the Gaussian distribution is therefore not a refinement of the Gaussian distribution, but rather an approximated distribution when this limit has not been reached.
Because the joint probability density of the envelope of the surface elevation $\zeta$ and its Hilbert transform $\hat{\zeta}$ can be well approximated by the expansion of third-order cumulants in terms of Hermite polynomials, \citet{Tayfun1994} showed that it can lead to the skewness $\mu_{3}$ correction in deep water:
\begin{equation}
f(\zeta, \hat{\zeta}) = \frac{e^{-\frac{1}{2}(\zeta^{2}+\hat{\zeta}^{2})}}{2\pi} \left[  1 + \frac{\mu_{3}}{6}  \zeta (\zeta^{2} +\hat{\zeta}^{2} - 4) \right] \quad .
\end{equation}
Integrating the joint density over the envelope $R=(\zeta^{2}+\hat{\zeta}^{2})^{1/2}$ with respective pair $(\zeta , \hat{\zeta}) = (R \cos{\phi} , R \sin{\phi})$, one finds the phase distribution in deep water \citep{Tayfun1994}:
\begin{equation}
f(\phi) = \frac{1}{2\pi} \left[  1 - \frac{\mu_{3}}{6} \sqrt{\frac{\pi}{2}} \cos{\phi} \right] \quad .
\label{eq:Tayfun}
\end{equation}
In deep water, we typically have $\mu_{3} \leqslant 0.6$ \citep{Tayfun2006,Tayfun2008}, such that the peak of the phase distribution does not exceed the uniform distribution $2\pi f(\phi)$ in 10\%. By comparing eq.~(\ref{eq:disthyp}) with eq.~(\ref{eq:Tayfun}) and the inequality of eq. 14 of \citet{Tayfun2006}, we find the bound:
\begin{equation}
\Xi_{1}  \geqslant  \frac{5}{6} (1+\nu^{2}) \left[ 1 + \mathcal{O}(\varepsilon)  \right] \gtrsim 1 \quad .
\label{eq:Tayfunbound}
\end{equation}
In intermediate and shallow waters, especially in unsteady conditions similar to the experiments of \citet{Trulsen2020}, this departure from the standard uniform distribution is expected to be much larger, see for instance figure 13 of \citet{Bitner1980}. However, approximations of Gram-Charlier series are reduced to either skewness or kurtosis effects, see \citet{Janssen2006a} for the latter. Nevertheless, the full computation of a Gram-Charlier series is dependent on both moments of the surface elevation \citep{Mori2002b}. Together with the explicit effect of bandwidth $\nu$ \citep{Higgins1975} from eq. 20 of \citet{Tayfun2006}, the impact of both skewness and kurtosis as computed in eq. 50 of \citet{Tayfun1989} reads:
\begin{eqnarray}
\nonumber
2\pi f(\phi) &=& 1 - \frac{\sqrt{\pi}}{4} ka \big(1- \nu \sqrt{2} + \nu^{2} \big) \cos{\phi} 
\\
&{}& + \, \big(\nu \sqrt{2} - \nu^{2} \big) (ka)^{2} \cos{(2\phi)} \, .
\end{eqnarray}
For a broad-banded sea state with typical JONSWAP spectrum of peakedness parameter $\gamma = 3.3$ the bandwidth is of the order of $\nu \sim 1/2$. By means of the relation between skewness and steepness \citep{Tayfun1989}, I find:
\begin{equation}
2\pi f(\phi) \approx 1 - \frac{\mu_{3}}{6} \sqrt{\frac{\pi}{2}} \cos{\phi} + \frac{9 (ka)^{2}}{20} \cos{(2\phi)} \quad .
\end{equation}
Furthermore, since eqs. 19-22 of \citet{Mori1998} show that the kurtosis is computed as $\mu_{4} = 48 (ka)^{2}$ in deep water (see appendix A of \citet{Mendes2023b}), I further simplify the phase distribution as follows:
\begin{equation}
2\pi f(\phi) \approx 1 - \frac{\mu_{3}}{6} \sqrt{\frac{\pi}{2}} \cos{\phi} + \frac{ \mu_{3}^{2}}{60} \cos{(2\phi)} \quad , 
\end{equation}
which adds another 0.4\% deviation to the maximum of 10\% of the first term containing $\cos{\phi}$. Hence, the approximation of eq.~(\ref{eq:Tayfun}) is justified in deep water. In intermediate water, the importance of the second term can increase by fivefold due to $\mu_{3} \sim 1$. In order to express the phase distribution in terms of relative water depth and wave steepness, I adopt the finite-depth formula of eq. 22 of \citet{Mori1998}. Taking into account the ratio between the coefficient in eq.~(\ref{eq:zetairregular}) and those of \citet{Mori1998} amounts to $\sqrt{\Tilde{\chi}_{1}}/(D_{1}+D_{2}) \sim 2$, the phase distribution fully reads:
\begin{eqnarray}
\nonumber
2\pi f(\phi) &\approx &   1 - \left( \frac{\pi \varepsilon \mathfrak{S}}{6}  \sqrt{\tilde{\chi}_{1}}  \right) \cos{\phi} 
\\
&{}& + \frac{6}{5\pi} \left( \frac{\pi \varepsilon  \mathfrak{S}}{6}  \sqrt{\tilde{\chi}_{1}}  \right)^{2}  \cos{(2\phi)}  ,
\end{eqnarray}
where the vertical asymmetry between crests and troughs $1 \leqslant \mathfrak{S} = 2a/H \leqslant 2$ is added to correct an otherwise symmetrical approach from the beginning. In comparison with eq.~(\ref{eq:disthyp}), the above solution is equivalent of finding $\Xi_{1} = 3/2\mathfrak{S} \approx 1.25$, upholding the lower bound in eq.~(\ref{eq:Tayfunbound}).
With the exact distribution for random phases in hand, I compute the change in mean water level:
\begin{widetext}
\begin{eqnarray}
\nonumber
\mu_{1}  &=&  \int_{0}^{2\pi} \frac{ d\phi \,      }{2\pi \sigma  \sqrt{\mathscr{E}}} \left[ \cos{\phi} + \frac{\pi \varepsilon \, \mathfrak{S}}{4} \sqrt{\tilde{\chi}_{1}} \cos{(2\phi) } \right] \left[  1 -  \frac{\pi \varepsilon \, \mathfrak{S} }{6}   \sqrt{\tilde{\chi}_{1}}  \cos{\phi} +  \frac{\pi \varepsilon^{2} \, \mathfrak{S}^{2}}{30}  \tilde{\chi}_{1}  \cos{(2\phi)}  \right] \,\, ,
\\
&=& \frac{\pi \varepsilon \mathfrak{S} \sqrt{2\tilde{\chi}_{1}} (\pi^{2}\varepsilon^{2} \mathfrak{S}^{2} \tilde{\chi}_{1}-20)}{ 240 \sqrt{ 1 + \frac{\pi^{2} \varepsilon^{2} \mathfrak{S}^{2}}{32} \left(  \tilde{\chi}_{1} + \chi_{1}\right)}   \sqrt{   1 + \frac{\pi^{2} \varepsilon^{2} \mathfrak{S}^{2} }{81} \tilde{\chi}_{1} + \frac{\pi^{4}\varepsilon^{4}\mathfrak{S}^{4} }{2300} \tilde{\chi}_{1}^{2} - \frac{\pi^{6}\varepsilon^{6}\mathfrak{S}^{6} }{284,000} \tilde{\chi}_{1}^{3} } } \equiv \frac{ 4 \mathbb{E}[ \zeta ] }{ H_{s} } \, .
\label{eq:Tayfun1}
\end{eqnarray}
\end{widetext}
\begin{figure*}
\minipage{0.35\textwidth}
    \includegraphics[scale=0.5]{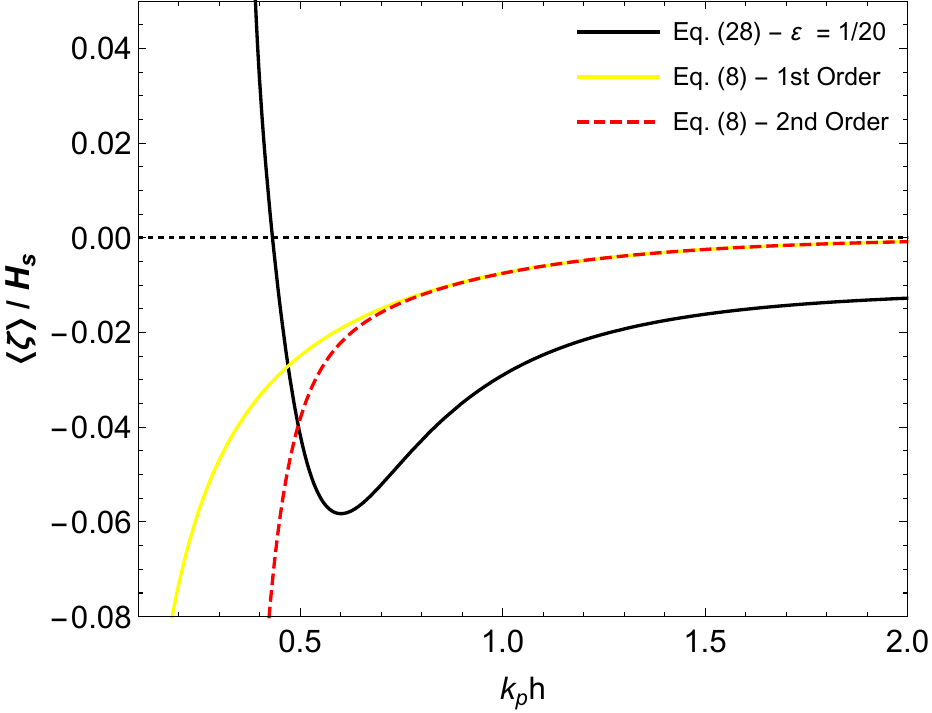}
\endminipage
\hfill
\minipage{0.5\textwidth}
    \includegraphics[scale=0.56]{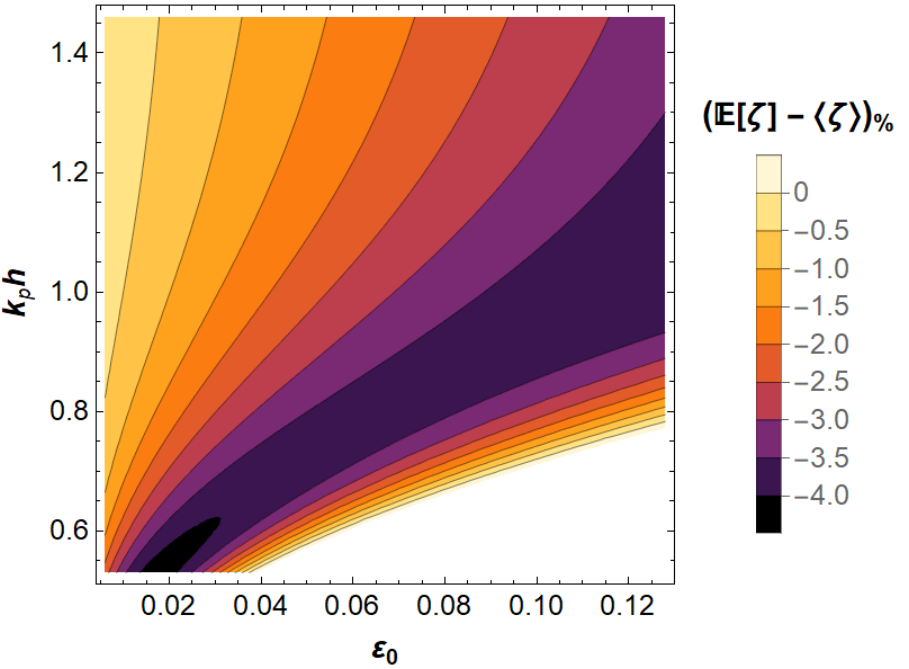}
\endminipage
\caption{(a) Set-down/up normalized by significant wave height as a function of water depth as computed from the non-ergodic formula eq.~(\ref{eq:Tayfun1}) and the classical approach of eq.~(\ref{eq:setdownRS}) adapted to irregular waves in eq.~(\ref{eq:setdownRSirregular}) for a fixed steepness $\varepsilon = 1/20$. (b) Excess set-down in percentage of $H_{s}$ by the non-ergodic computation as compared to the classical counterpart with varying pre-shoal steepness $\varepsilon_{0}$ and relative water depth.}
\label{fig:moment122}
\end{figure*}
In \jfm{figure} \ref{fig:moment11} the \xx{theoretical} evolution of the normalized \xx{mean water level} is \xx{displayed}, decreasing evermore towards shallow water until it reaches its \xx{global minimum at the surf zone}. Then, the mean water level starts to increase and reaches the \xx{plunging point at $\mu_{1}=0$} and quickly increases to a normalized set-up, as expected from observation \citep{daSilva2020}. \jfm{Figure} \ref{fig:moment11} \xx{also points to the fact that although increasing the pre-shoal steepness leads to an earlier peak in set-down, the magnitude of this peak seems to vary weakly with the steepness. This is in qualitative} agreement with the \xx{linear term of the} set-down formula of eq.~(\ref{eq:setdownRS}), \xx{because the latter converges to $-a^{2}/4h$ in the surf zone. Therefore, the new formula recovers both sides of the piecewise theory of \citet{Higgins1962}, and being continuous explains theoretically when and how fast the transition between set-down and set-up occurs}.

\begin{figure}
    \centering
    \includegraphics[scale=0.56]{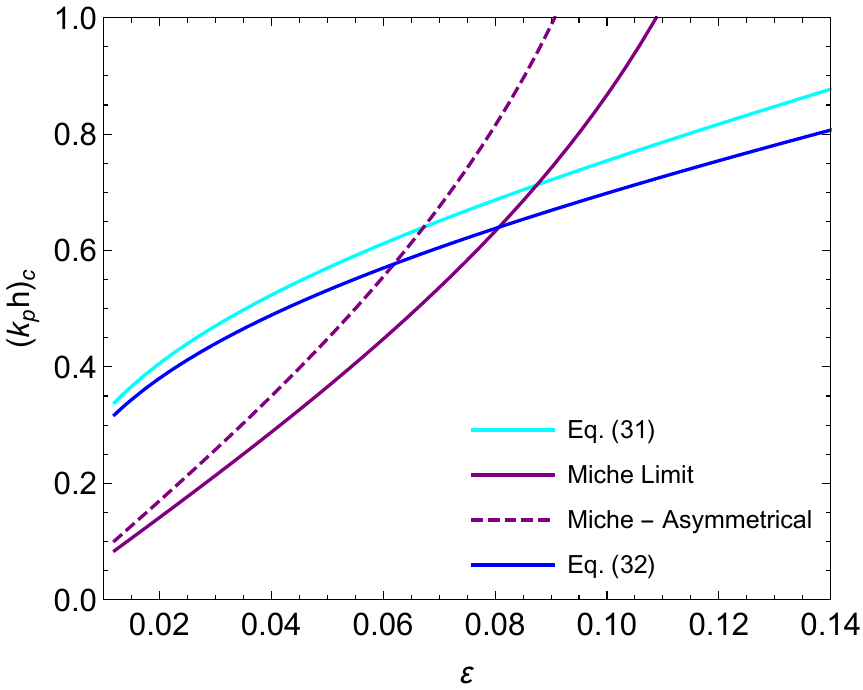}
    \caption{Critical values (breaking, plunging) for the relative water depth as a function of curves of fixed steepness according to the present theory and steepness-limited wave breaking \citep{Miche1944} with and without vertical wave asymmetry.}
    \label{fig:khcritical}
\end{figure}
In order to compare the set-down computed from the random phase distribution with the radiation stress \xx{theory}, eq.~(\ref{eq:setdownRS}) is rewritten for irregular waves by using the same transformation from eq.~(\ref{eq:zetairregular}):
\begin{equation}
\frac{\langle \zeta \rangle}{\sigma} \approx \frac{4\langle \zeta \rangle}{H_{s}} = - \frac{\pi \varepsilon \mathfrak{S} }{\sqrt{2}\sinh{(2.2k_{p}h)}} \left[ 1 + \frac{9 \pi^{2} \varepsilon^{2} \mathfrak{S}^{2} }{16 \sinh^{6}{(1.1k_{p}h)}}    \right] \, .
\label{eq:setdownRSirregular}
\end{equation}
Thus, \jfm{figure} \ref{fig:moment122} compares the non-ergodic set-down from the ensemble average with the time average from eq.~(\ref{eq:setdownRSirregular}).
The set-down computation based on the radiation stress of waves over a flat bottom underpredicts the result of the present model, and this is expected because the latter is based on the $\Gamma$ model that is valid for relatively steep slopes. This is in agreement with the fact that \citet{Higgins1962} underpredicts the set-down in the experiments performed by \citet{Saville1961} for steep slopes \citep{Higgins1963,Ehrenmark1994,Hsu2006b}.

In addition, it seems that the point in which wave breaking occurs, which in the present model first moment plot corresponds to the transition between set-down and set-up, is well predicted by the mean. To find the breaking point one uses $\varepsilon \leqslant \tanh{(kh)}/7 $ and thus find $ kh \geqslant \tanh^{-1} (7\varepsilon)$ according to \citet{Miche1944}. Factoring out $\sqrt{\mathscr{E}}$ which is of the order of $\mathcal{O}(1)$, The breaking point can be obtained by setting $d\mu_{1}/d\tilde{\chi}_{1} = 0$, finding to leading order:
\begin{equation}
\tilde{\chi}_{1}  \approx \frac{12 }{\pi^{2} \varepsilon^{2} \mathfrak{S}^{2} } \left[ \sqrt{1 + \frac{10\pi}{9} } - 1 \right] \approx \frac{40 }{3\pi^{2} \varepsilon^{2} \mathfrak{S}^{2} }  \quad .
\end{equation}
Following the definition of the trigonometric coefficient $\tilde{\chi}_{1} $ in eqs.~(\ref{eq:zetairregular}-\ref{eq:Gamma}), I solve the cubic equation in $\tanh{k_{p}h}$ and find the critical relative depth for $\mathfrak{S} \approx 1.2$:
\begin{eqnarray}
\nonumber
k_{p}h &\approx& - \tanh^{-1} \left[ \frac{\varepsilon }{3} - \frac{ (1 + i \sqrt{3} ) \varepsilon^{2} }{ 6\psi^{1/3}  }    - \frac{ (1 - i \sqrt{3} ) \psi^{1/3}}{ 6 }  \right] \, , 
\\
\psi &=&  \varepsilon^{3} + \frac{81 \varepsilon}{2} \left(  \sqrt{ 1 - \frac{4 \varepsilon^{2}}{81} }  -1  \right) \, .
\end{eqnarray}
I can also obtain the location of the plunging/spilling point by solving $\mu_{1}(\varepsilon, (k_{p}h)_{c}) = 0$ which implies $\tilde{\chi}_{1}  = 20/\pi^{2} \varepsilon^{2} \mathfrak{S}^{2}$:
\begin{eqnarray}
\nonumber
k_{p}h &=& - \tanh^{-1} \left[ \frac{5\varepsilon }{18} - \frac{25 (1 + i \sqrt{3} ) \varepsilon^{2} }{ 36\tilde{\psi}^{1/3}  }    - \frac{ (1 - i \sqrt{3} ) \tilde{\psi}^{1/3}}{ 36 }  \right]  ,
\\
\tilde{\psi} &=& 125 \, \varepsilon^{3} + 7290 \, \varepsilon \left(  \sqrt{ 1 - \frac{25 \varepsilon^{2}}{729} }  -1  \right) \, .
\end{eqnarray}
\jfm{Figure} \ref{fig:khcritical} compares steepness-limited wave breaking critical points with the breaking and plunging points as estimated by the present theory, showing qualitative agreement. However, the model shows that depth-limited wave breaking turns the transition between set-down and set-up earlier than inferred from \citet{Miche1944} at low wave steepness.

\section{Slope Effect}

To fully compute \xx{the effect of steep slopes on the mean  water level, one must start with its effect on} the derivative of the velocity potential. The algebra can be split into slope-dependent and slope-independent \xx{terms}:
\begin{equation}
u =  \frac{\partial \Phi}{\partial x} = u_{0} + \Delta u (\nabla h) \quad ; \quad w =  \frac{\partial \Phi}{\partial z} \quad .
\end{equation}
Hence, the slope-dependent term of the horizontal velocity component will lead to a modification in eq.~(\ref{eq:Bernoulli}):
\begin{figure}
\centering
    \includegraphics[scale=0.56]{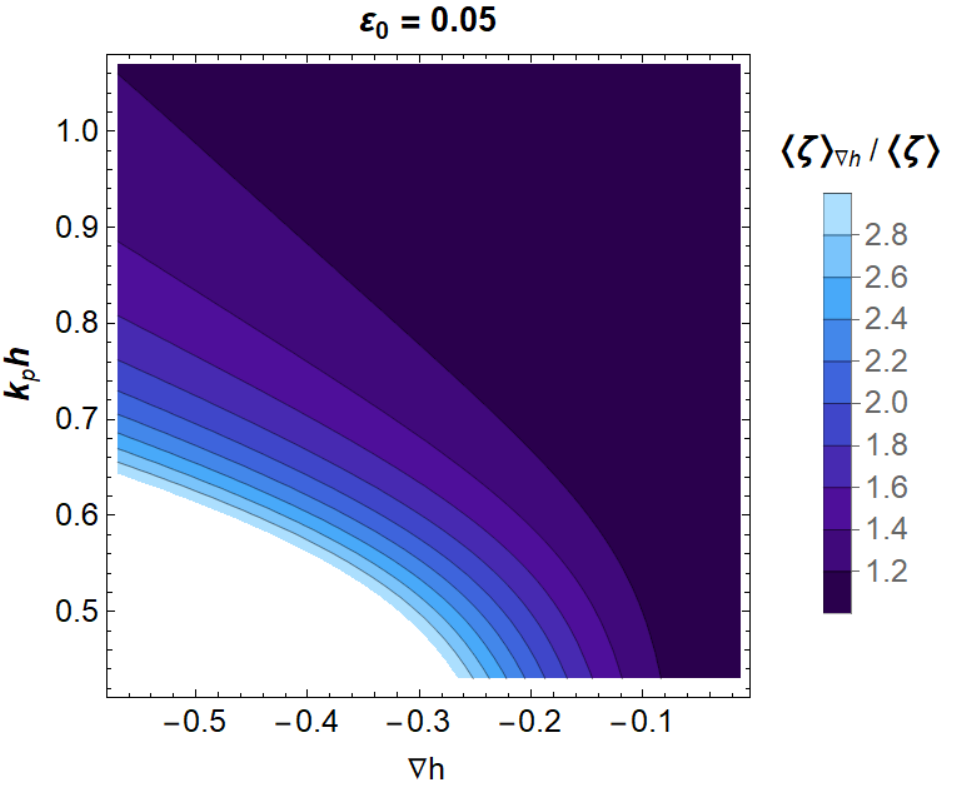}
\caption{Amplification in the classical set-down formula due to the slope effect of eq.~(\ref{eq:setdownRSslope}).}
\label{fig:moment12}
\end{figure}
\begin{eqnarray}
\nonumber
\langle \zeta \rangle_{_{\nabla h}} &=& -\frac{\Big\langle (u_{0} + \Delta u )^{2} - w^{2} \Big\rangle}{2g}  
\\
&=& \langle \zeta \rangle -\frac{\Big\langle 2 u_{0} \Delta u + (\Delta u )^{2} \Big\rangle}{2g}  ,
\end{eqnarray}
with the term $\langle \zeta \rangle$ having been already obtained in eq.~(\ref{eq:setdownRS}). The complementary part of the horizontal velocity is computed by taking derivatives $\partial/\partial x$ on the hyperbolic functions in eq.~(\ref{eq:U}) since $\partial \theta / \partial x = \partial \Lambda / \partial x = k \nabla h$:
\begin{eqnarray}
\frac{k \Delta u }{a\omega} =   \sin{\phi} \, \pd{}{x} \left[ \frac{\cosh{\theta} }{\sinh{\Lambda}}  \right]  +  \frac{3ka }{8}  \sin{(2\phi)}  \,  \pd{}{x} \left[ \frac{\cosh{(2\theta)} }{\sinh^{4}{\Lambda}}  \right] ,  
\label{eq:U2}    
\end{eqnarray}
\xx{and thus one can obtain}:
\begin{eqnarray}
\nonumber
\Delta u &=& \frac{a \omega \nabla h}{\sinh^{2}{\Lambda}}  \left\{ \mathscr{H}_{1} \sin{\phi}    + \left(  \frac{3ka}{4}  \right) \frac{\mathscr{H}_{2} \sin{(2\phi)} }{\sinh^{4}{\Lambda}} \right\} ,
\\
\nonumber
\mathscr{H}_{1} &=&  \sinh{\theta} \sinh{\Lambda} - \cosh{\theta} \cosh{\Lambda} \,\, ,
\\
\mathscr{H}_{2} &=&  \sinh{(2\theta)} \sinh^{2}{\Lambda} - \cosh{(2\theta)} \sinh{(2\Lambda)} \,\, .
\end{eqnarray}
Because of eq.~(\ref{eq:sincos}), the time average of $u_{0} \Delta u$ vanishes. \xx{Although} the hyperbolic coefficients appearing in $(\Delta u)^{2}$ \xx{are bulky and can not be simplified} I may take the limit at $z = 0$ for these coefficients \xx{due to} $\langle u_{0}^{2}-w^{2} \rangle / 2g \sim \langle (\Delta u)^{2} \rangle / 2g < 0.1$. In that case, it is straightforward to show that $\lim_{z \rightarrow 0} \mathscr{H}_{1} = 1$ and $\lim_{z \rightarrow 0} \mathscr{H}_{2} = -2\sinh^{4}{\Lambda}/\tanh^{3}{\Lambda}$. Therefore, I may approximate:
\begin{widetext}
\begin{eqnarray}
\nonumber
\left\langle  \frac{(\Delta u )^{2}}{2g}  \right\rangle &=&  \frac{}{} \frac{a^{2}\omega^{2} (\nabla h )^{2}}{2g\sinh^{4}{\Lambda}} 
 \left\langle
\mathscr{H}_{1}^{2} \sin^{2}{\phi} +  \left(  \frac{3ka}{4}  \right)^{2} \frac{\mathscr{H}_{2}^{2} \sin^{2}{(2\phi)} }{\sinh^{8}{\Lambda}}
\right\rangle 
= \frac{a^{2}}{2g} \cdot \frac{gk \tanh{\Lambda}\, (\nabla h )^{2}}{2\sinh^{4}{\Lambda}} \left[
\mathscr{H}_{1}^{2}  +  \left(  \frac{3ka}{4}  \right)^{2} \frac{\mathscr{H}_{2}^{2} }{\sinh^{8}{\Lambda}} 
\right]   \, ,
\\
&=& \frac{ka^{2}}{2\sinh{(2\Lambda)}} \cdot \frac{ (\nabla h )^{2}}{\sinh^{2}{\Lambda}} \left[
\mathscr{H}_{1}^{2}  +  \left(  \frac{3ka}{4}  \right)^{2} \frac{\mathscr{H}_{2}^{2} }{\sinh^{8}{\Lambda}} 
\right] \approx -\frac{ka^{2}}{2\sinh{(2\Lambda)}} \cdot \frac{ (\nabla h )^{2}}{\sinh^{2}{\Lambda}} \left[
1 +   \frac{9(ka)^{2}}{4\tanh^{6}{\Lambda}} \right] \quad .
\end{eqnarray}
\end{widetext}
Accordingly, up to second order in steepness $ka$ and under the effect of an arbitrary slope $\nabla h$ without curvature $(\nabla^{2}h = 0)$, the wave-driven set-down is fully computed at last:
\begin{widetext}
\begin{equation}
\langle \zeta \rangle_{_{\nabla h}} \approx - \frac{ka^{2}}{2\sinh{(2kh)}} \left\{ 1 + \frac{9 (ka)^{2}}{16 \sinh^{6}{kh}} +   \frac{ (\nabla h )^{2}}{\sinh^{2}{kh}} \left[
1 +   \frac{9(ka)^{2}}{4\tanh^{6}{kh}} 
\right]   \right\} \quad .
\label{eq:setdownRSslope}
\end{equation}
\end{widetext}
\xx{In \jfm{figure} \ref{fig:moment12} the disparity between mild slope and steep slope mean water levels is displayed.} Taking into account the magnitude of the maximal correction due to the terms proportional to $(ka)^{2}/(kh)^{6}$ approaching the Ursell limit in shallow water, I can further simplify the slope-dependent set-down:
\begin{figure}
    \centering
    \includegraphics[scale=0.48]{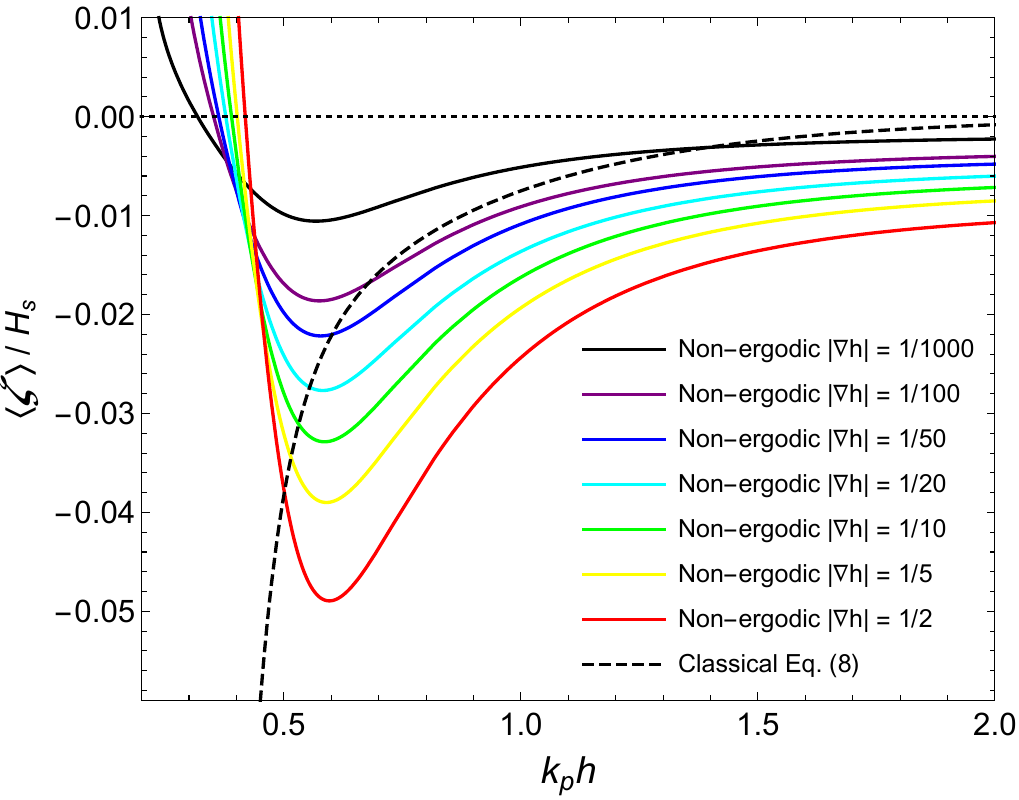}
    \caption{Slope dependence of the non-ergodic computation for the set-down/up.}
    \label{fig:setdownslopes}
\end{figure}
\begin{eqnarray}
\nonumber
\langle \zeta \rangle_{_{\nabla h}} &\approx& - \frac{5ka^{2}}{9\sinh{(2kh)}} \left[ 1 + \frac{4}{3}   \frac{ (\nabla h )^{2}}{\sinh^{2}{kh}} \right] 
\\
&\approx& \langle \zeta \rangle \cdot \frac{10}{9} \left[ 1 + \frac{4}{3}   \frac{ (\nabla h )^{2}}{\sinh^{2}{kh}} \right] \quad .
\label{eq:setdownRSslope2}
\end{eqnarray}
The above slope correction is the first of the kind to be found analytically through the classical textbook methodology, albeit the dependence on the slope has been found previously through numerical simulations \citep{Ehrenmark1994} and experiments \citep{Holman1985}. Although \citet{Hsu2006b} have also found a dependence on $(\nabla h)^{2}$ for the shoaling-induced set-down, this comes from an \textit{a priori} expansion of the velocity potential in powers of $ka \cdot \nabla h$ and the formulae for the computation contains dozens of hyperbolic and trigonometric coefficients that make such a result computationally cumbersome. Notably, the present derivation does not apply to realistic reflective beaches, so the above formula is limited to cases with a negligible reflection coefficient $K_{R} \approx 0.1 \upxi^{2} $, hence with small surf similarity $\upxi \sim \nabla h /  \sqrt{\varepsilon} \lesssim 1 $ \citep{Holthuijsen2007}. This limitation inhibits an otherwise divergence in the set-down due to a step $\nabla h \rightarrow \infty$. On the other hand, one may examine the maximum set-down beyond second-order theory in the limit when $\sinh{kh} \approx \tanh{kh} \approx kh$:
\begin{eqnarray}
\frac{\langle \zeta \rangle_{_{\nabla h}}}{H_{s}} \geqslant - \frac{H_{s}}{32h} \left[ 1  + 10 \, (\nabla h )^{2} \left\{  1 + \frac{9}{2} \left( \frac{ H_{s}}{h} \right)^{2} \right\} \right]   \, .
\end{eqnarray}
Therefore, the maximum set-down using linear theory is amplified by a percentage of $ (50\nabla h )^{2} \, \%$ due to steep slopes, which for negligible shoal reflection lies in the range $25\% - 250\%$ for $1/10 < |\nabla h| < 1/3$. This amplification magnitude is at par with two to threefold larger set-down than predicted by \citet{Higgins1962} as compared to experiments in \citet{Saville1961}. However, the set-up can not be computed from eq.~(\ref{eq:setdownRSslope2}) and no continuous formulation can be extended. In the next section, I compute and show that the same shape of the set-down zone can be obtained, although it is more complex algebraically but with the advantage of computing the mean water level continuously from set-down up to set-up.


\subsection{Non-ergodic Approach}

\begin{figure}
\centering
    \includegraphics[scale=0.6]{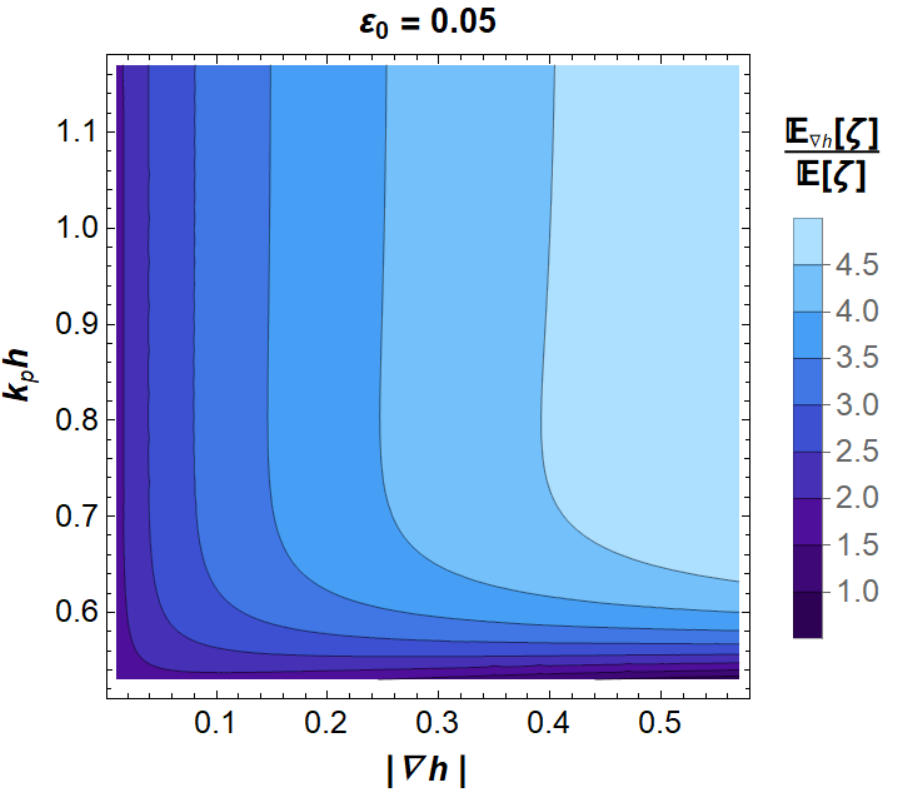}
\caption{Amplification in the mean water level formula due to the slope effect of eq.~(\ref{eq:nonergodicsetdown}) as compared to a fixed mild slope of $|\nabla h| = 1/1000$.}
\label{fig:moment13}
\end{figure}

In order to extract the slope dependence from the set-down, first one must generalize the random phase approach for eq.~(\ref{eq:randomphaseGamma}). As discussed in \citet{Mendes2023b}, the \xx{shoaling} slope effect induces a correction $ \sqrt{\nabla h}$ to the existing excess kurtosis of steep slopes, thereby decreasing the non-Gaussianity as the slope tends to zero.  Combination of the "uniform" slope-dependent phase distribution with the inhomogeneous slope-independent phase distribution of eq.~(\ref{eq:Tayfun1}) leads to \xx{(see \jfm{appendix} \ref{sec:appslope})}:
\begin{eqnarray}
\hspace{-0.3cm}
\nonumber
2\pi f_{_{\nabla h}}(\phi) &\approx&   1 - \left( \frac{\pi \varepsilon \, \mathfrak{S}\sqrt{\tilde{\chi}_{1}}}{6}    \right) \xx{|\nabla h|^{1/4}} \cos{\phi}
\\
&{}& + \frac{6 }{5\pi} \left( \frac{\pi \varepsilon \, \mathfrak{S} \sqrt{\tilde{\chi}_{1}}}{6}    \right)^{2} \xx{|\nabla h|^{1/2}}  \cos{(2\phi)}  \, .
\label{eq:fulldist}
\end{eqnarray}
Naturally, the generalized slope-dependent distribution recovers both inhomogeneous and uniform distributions over mild slopes. Such distribution allows us to estimate the effect of arbitrarily steep slopes on the set-down, as an alternative to the computation of the Bernoulli equation (or radiation stress) leading to eq.~(\ref{eq:setdownRSslope2}). As such, the set-down driven by the equivalent inhomogeneous random phase distribution of eq.~(\ref{eq:Tayfun1}) now reads:
\begin{widetext}
\begin{equation}
 \mu_{1} = \frac{\pi \varepsilon \mathfrak{S}  \sqrt{2 \tilde{\chi}_{1}}      (\pi^{2}\varepsilon^{2} \mathfrak{S}^{2} \tilde{\chi}_{1} \sqrt[4]{\nabla h} - 20) \sqrt[4]{\nabla h   } \left[ 1 + \frac{\pi^{2} \varepsilon^{2} \mathfrak{S}^{2}}{32} \left(  \tilde{\chi}_{1} + \chi_{1}\right) \right]^{-1/2} }{ 240  \sqrt{   1 + \frac{\pi^{2} \varepsilon^{2} \mathfrak{S}^{2}   }{720} \tilde{\chi}_{1}  ( 45 - 30 \sqrt[4]{\nabla h} - 6 \sqrt{\nabla h}  )  + \frac{\pi^{4}\varepsilon^{4}\mathfrak{S}^{4}  }{2300} \tilde{\chi}_{1}^{2} (\nabla h)^{3/4}  - \frac{\pi^{6}\varepsilon^{6}\mathfrak{S}^{6}  }{142,000} \tilde{\chi}_{1}^{3} \, \nabla h   } } \, ,
\label{eq:nonergodicsetdown}
\end{equation}
\end{widetext}
As observed in \jfm{figure} \ref{fig:setdownslopes}, as the slope becomes milder ($|\nabla h| \leqslant 1/100$) the non-ergodic set-down converges to the adiabatic classical set-down of \citet{Higgins1962}. Comparing it with the slope-dependent classical set-down computation introduced in eq.~(\ref{eq:setdownRSslope}), the magnitude of the global minimum is similar, but the shape of the set-down evolution is different. Moreover, the slope-dependent classical set-down only deviates from the adiabatic one near wave breaking, thus a shortcoming. Naturally, the present formula is superior as it is also capable of computing the set-up, in particular also describing its dependence on the slope. Lastly, \jfm{figure} \ref{fig:moment13} describes how a steep slope amplifies the set-down, but this growth is significant only in a narrow range of relative water depth $0.6 \leqslant k_{p}h \leqslant 0.9$.

\section{Conclusion}

In this work the radiation stress-led computation of the set-down has been extended to include the effect of slope. Furthermore, I have shown that statistical reasoning removes the difficulties in dealing with wave dissipation for the computation of the mean water level when calculated with a non-uniform distribution of random phases, being general for steep slopes as well. By comparing the two formulations, it can be explained why the classical calculation underpredicted the set-down over steep slopes, being recovered by the new slope-dependent model for slopes milder than 1/100. However, the radiation stress can still not lead to a continuous formulation between set-down and set-up, whereas the stochastic approach succeeds in this matter. Despite the usage of a simple model for the distribution of random phases, this work demonstrates that wave statistics are useful for computing primary variables of the water wave solutions, paving the way for the realization that physically visible effects can be handled by the intangible randomness measures of a sea state.

\textbf{Declaration of Interests}. The author reports no conflict of interest.

\appendix

\section{Rayleigh Distribution}\label{sec:Rayapp}

A straightforward proof that a Rayleigh distribution will emerge from the magnitude of a random vector reveals the uniform distribution of random phases: given two random variables of unknown form $X = \sum_{i} x_{i}$ and $Y = \sum_{i} y_{i}$, mutually independent and identically distributed by a Gaussian probability density due to the \textit{central limit theorem}, its joint probability density is:
\begin{equation}
f_{XY}=  \frac{1}{2\pi \sigma^{2}} e^{-(X^{2}+Y^{2})/2\sigma^{2}} \quad .
\end{equation}
Choosing auxiliary random variables $X = R \, \cos {\Omega}$ and $Y = R\, \sin {\Omega}$, equivalent of the surface elevation and its Hilbert transform \citep{Mori2002}, the Jacobian reads:
\begin{equation}
\left| \frac{\partial (X,Y)}{\partial (R,\Omega)} \right| = 
\begin{vmatrix}
\cos {\Omega} & \sin {\Omega} \\
- R \, \sin {\Omega} & R \, \cos {\Omega}
\end{vmatrix}
= R\quad .
\end{equation}
The marginal exceedance distribution of the envelope $R$ \xx{if} phases are uniformly distributed $(f_{\Omega}=1/2\pi)$ \xx{becomes}:
\begin{equation}
\mathbb{P}_{R} = \int_{R}^{+\infty} \int_{0}^{2\pi} \frac{e^{-R^{\ast \, 2}/2\sigma^{2}}}{2\pi \sigma^{2}} R^{\ast} \, dR^{\ast} \, d\Omega = e^{-R^{2}/2\sigma^{2}} \,\, , 
\end{equation}
\xx{whose derivative is the Rayleigh probability density:}
\begin{equation}
f_{R} = \frac{R}{\sigma^{2}} e^{-R^{2}/2\sigma^{2}} \quad .
\end{equation}

\section{\xx{Slope Effect on Kurtosis:\\ Taylor Expansion}}\label{sec:appslope}

\begin{figure*}
\centering
    \includegraphics[scale=0.6]{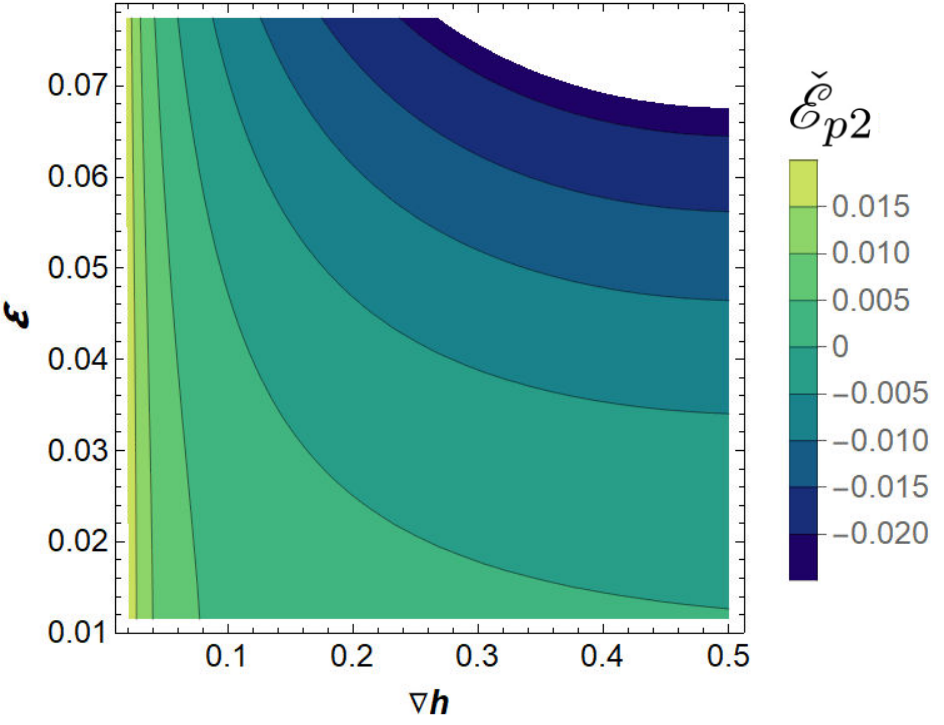}
\caption{Net change in spectral potential energy corrected by boundary terms as described in eq.~(\ref{eq:ep2}).}
\label{fig:ep2}
\end{figure*}
\xx{Here I demonstrate why the effect on slope is approximately factored out as $\sqrt{\nabla h}$ in the excess kurtosis. From section 3 of \citet{Mendes2023b} an effective approximation for the excess kurtosis has been experimentally validated for waves traveling at steep slopes:}
\begin{eqnarray}
\xx{ \mu_{4} (\Gamma) \approx \frac{1}{9} \left[ e^{ 8 \left( 1 - \frac{ 1 }{ \mathfrak{S}^{2}\Gamma } \right) }- 1 \right]     \quad .
}
\label{eq:kurtslope}
\end{eqnarray}
\xx{Without loss of generality, I simply the process of performing a Taylor expansion by setting a representative value of the vertical asymmetry that varies slowly \citep{Mendes2023b}. For waves with high steepness (in the range of second-order theory) similar to the conditions in \jfm{figure} \ref{fig:moment122} ($\varepsilon \lesssim 1/20$) or the experiments reviewed in \citet{Chabchoub2023} I can approximate $\mathfrak{S}\Gamma \approx \Gamma^{6}$ because $\Gamma \lesssim 1.05$, see eq. 3.26 of \citet{Mendes2021b}. Furthermore, including the slope effect of eq. 29 of \citet{Mendes2022b}, the non-homogeneous spectral correction formulated in \jfm{section} \ref{sec:setdown} can also be approximated as:}
\begin{eqnarray}
\xx{ \Gamma \approx 1 + \left( \frac{\pi \varepsilon}{4} \right)^{2}  \left( \frac{ \tilde{\chi}_{1} - \chi_{1} }{2}\right) - \check{\mathscr{E}}_{p2} \quad  , }
\end{eqnarray}
\xx{which at the relative water depth leading to the highest amplification ($k_{p}h = 0.5$) leads to:}
\begin{figure}
\centering
    \includegraphics[scale=0.56]{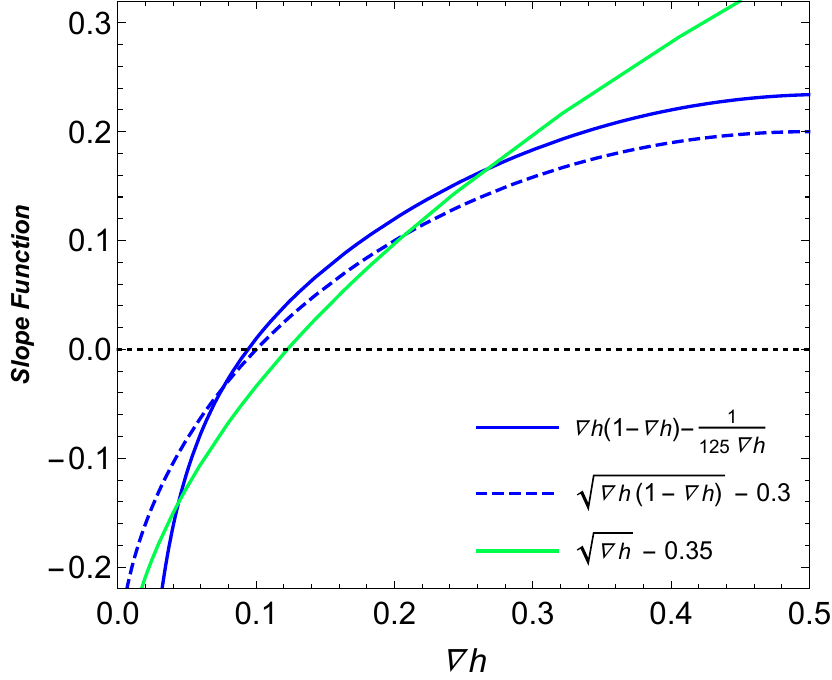}
\caption{Approximations for the boundary-adjusted slope function appearing in eq.~(\ref{eq:ep2}).}
\label{fig:approxslope}
\end{figure}
\begin{eqnarray}
\xx{ \Gamma \approx 1 + 2 \pi^{2}  \varepsilon^{2} - \check{\mathscr{E}}_{p2}   \quad . }
\end{eqnarray}
\xx{Here, $\check{\mathscr{E}}_{p2}$ is the net change in spectral potential energy due to the slope effect on the mean water level adjusted by a boundary term proportional to $(\nabla h)^{-1}$, being expressed as a function of the slope magnitude $\nabla h$ in the case of a slope as (also at $k_{p}h = 0.5$ and with pre-shoal water depth $k_{p0}h_{0} = \pi$, see \jfm{figure} \ref{fig:ep2}):}
\begin{equation}
\xx{
\check{\mathscr{E}}_{p2} \approx 20\varepsilon^{2} \left[ -  \nabla h \left( 1 - \nabla h    \right) +  \frac{ 1 }{125 \nabla h }   \right] .
}
\label{eq:ep2}
\end{equation}
\xx{As shown in \jfm{figure} \ref{fig:approxslope}, I can rewrite the net potential energy by numerically simplifying its closed-form:}
\begin{equation}
\xx{
-\check{\mathscr{E}}_{p2} \approx \varepsilon^{2} \left( 20 \sqrt{ \nabla h } -7   \right) .
}
\end{equation}
\xx{Note that this approximation clearly shows that the slope effect starts to saturate at slopes about twice the critical point $(\nabla h)_{c} = (7/20)^{2} \sim 1/8$.  Then, the exponent can be rewritten as,}
\begin{eqnarray}
\nonumber
 1 - \frac{ 1 }{ \mathfrak{S}^{2}\Gamma }    &\approx & 1 - \left[ 1 + \varepsilon^{2}  \big( 2 \pi^{2} - 7 +  20 \sqrt{ \nabla h } \big)   \right]^{-6}  \, ,
 \\
 \nonumber
&\approx & 1 - \left[ 1 - 6\varepsilon^{2}  \big( 2 \pi^{2} - 7 +  20 \sqrt{ \nabla h } \big)   \right]  \, ,
\\
&\approx &  12 \pi^{2} \varepsilon^{2} + 6\varepsilon^{2} \big(   20 \sqrt{ \nabla h } - 7 \big)    \, .
\end{eqnarray}
\xx{Comparing to eq.~(\ref{eq:kurtslope}), the leading term of the excess kurtosis is of the order,}
\begin{equation}
\mu_{4 \, , \, 0} = \frac{1}{9} \left[ e^{ 96 \pi^{2} \varepsilon^{2}}- 1 \right] \xrightarrow{\varepsilon \rightarrow 1/20}  1 \, ,
\end{equation}
\xx{as observed in well-known experiments \citep{Trulsen2020}. Note however that the above formula works for small amplitude waves, and a much larger value of steepness will require corrections up to 3rd and 4th orders. Moreover, because $e^{ 96 \pi^{2} \varepsilon^{2}} \gg 1$ I can use:}
\begin{eqnarray}
\mu_{4} &=& \frac{ e^{ -48 \check{\mathscr{E}}_{p2} } }{9} \left[ e^{ 96 \pi^{2} \varepsilon^{2}}- 1 \right] + \frac{1}{9} \left( e^{ -48 \check{\mathscr{E}}_{p2}} -1 \right) \, ,
\\
&=& \mu_{4 \, , \, 0} \cdot   e^{ -48 \check{\mathscr{E}}_{p2} } + \mathcal{O}(10\check{\mathscr{E}}_{p2})  \quad .
\end{eqnarray}
\xx{The second term is at least one order of magnitude smaller than the first, and I can neglect it. I compute the exponential of the potential energy variation, obtaining an expansion for the assumed sea conditions ($400 \varepsilon^{2} \sim 1$):}
\begin{eqnarray}
\nonumber
e^{ -48 \check{\mathscr{E}}_{p2}} &\approx & 1 - 48 \check{\mathscr{E}}_{p2}  \approx 1 + 48 \varepsilon^{2} \big(   20 \sqrt{ \nabla h } - 7 \big) ,
\\
\nonumber
&\approx & (1 - 336\varepsilon^{2} ) + 960 \varepsilon^{2}  \sqrt{ \nabla h }  \,\, ,
\\
&\sim & 1000 \, \varepsilon^{2}  \sqrt{ \nabla h } \lesssim 2 \sqrt{ \nabla h }  \quad .
\end{eqnarray}
\xx{Expanding the exponential $20 \leqslant \mu_{4 \, , \, 0} / \pi^{2} \varepsilon^{2} \leqslant 40 $ (see \jfm{figure} \ref{fig:Taylormu0}), I conclude the proof for the second-order small amplitude wave theory in steepness ($\varepsilon \leqslant 1/20$):}
\begin{equation}
\mu_{4} \lesssim 80 \pi^{2} \varepsilon^{2} \sqrt{\nabla h} \quad . 
\end{equation}
\xx{It can be seen in \jfm{figure} \ref{fig:Taylormu0} that for low steepness one should probably use the coefficient lower bound of $40$ to compensate for the gap between solid and dashed curves, whereas for the bulk of the range in steepness the upper bound is the best approximation.}
\begin{figure}[tbh!]
\hspace{-0.5cm}
\centering
    \includegraphics[scale=0.66]{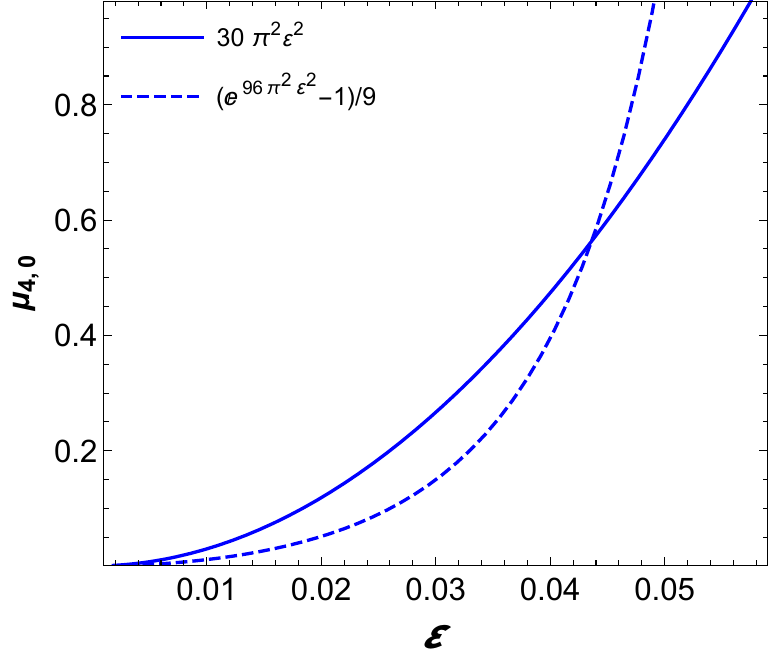}
\caption{Approximation for the slope-saturated excess kurtosis varying with steepness.}
\label{fig:Taylormu0}
\end{figure}

\bibliography{Maintext}

\end{document}